\documentclass{ifacconf}
\pdfoutput=1
\usepackage{lipsum} 
\usepackage{booktabs}
\usepackage{comment}
\usepackage{soul,color}
\usepackage[no comma]{optidef}
\usepackage{amsmath,amsfonts,amssymb}
\usepackage{graphicx}      
\usepackage[numbers]{natbib}  

\begin{document}
\begin{frontmatter}


\title{Safety-Driven Battery Charging: A Fisher Information-guided Adaptive MPC with Real-time Parameter Identification} 

\author[First]{Jorge Espin,} 
\author[First]{Yuichi Kajiura,} 
\author[First]{Dong Zhang}

\address[First]{School of Aerospace and Mechanical Engineering, The University of Oklahoma, Norman, OK 73019, USA (e-mail: [jorge.espin,~yuichi,~dzhang]@ou.edu)}

\begin{abstract}                
Lithium-ion (Li-ion) batteries are ubiquitous in modern energy storage systems, highlighting the critical need to comprehend and optimize their performance. Yet, battery models often exhibit poor parameter identifiability which hinders the development of effective battery management strategies and impacts their overall performance, longevity, and safety. This manuscript explores the integration of Fisher Information (FI) theory with Model Predictive Control (MPC) for battery charging. The study addresses the inherent hurdles in accurately estimating battery model parameters due to nonlinear dynamics and uncertainty. Our proposed method aims to ensure safe battery charging and enhance real-time parameter estimation capabilities by leveraging adaptive control strategies guided by FI metrics. Simulation results underscore the effectiveness of our approach in mitigating parameter identifiability issues,  offering promising solutions for improving the control of batteries during safe charging process.

\end{abstract}

\end{frontmatter}
\vspace{0cm}
\section{Introduction}

\vspace{-0.1cm}
In recent years, the widespread adoption of battery-powered technologies, spanning from electric vehicles to renewable energy storage systems, has underscored the critical importance of safe and efficient battery charging. Improper management during charging can lead to inefficient battery utilization, diminished battery safety, and premature battery degradation (\cite{komsiyska2021critical}). Therefore, there exists a pressing need to design advanced battery management systems to optimize charging processes. Among the various techniques available in the literature, model predictive control (MPC) stands out for its ability to leverage the underlying battery dynamics and operational constraints within an optimization framework to achieve desired objectives. This advanced control strategy has been validated by several studies. \cite{klein2011optimal} proposed a nonlinear model predictive control (NMPC) to rapidly charge batteries while ensuring safety throughout their lifespan. \cite{zou2018model} introduced a battery charging methodology to achieve an optimal balance between charging speed and battery State-of-Health (SoH). This approach involves developing a simplified physics-based battery model, which enables the design of a model-driven charging strategy. Additionally, \cite{pozzi2020optimal} outlined a sensitivity-based MPC with reduced computational cost suitable for battery packs comprising numerous cells.

However, a common limitation of these charging control methods is their high reliance on accurate battery models. 
Battery models, including empirical models and physics-based models, often struggle with parametric uncertainties caused by identifiability issues steaming from complex nonlinearities and inherent parameter correlations.  For instance, \cite{forman2012genetic} employed Fisher information (FI) metrics to assess parameter identifiability in electrochemical models, demonstrating that certain battery parameters cannot be identified using traditional charge and discharge protocols. In this context, identifiability refers to the ability to uniquely determine model parameters from available data, a problem exacerbated by limited and noisy experimental measurements (\cite{guillaume2019introductory}).

In response to poor battery parameter identifiability, researchers leveraged optimal experimental design (OED), a strategy aimed at designing experiments to maximize information about model parameters
by carefully selecting input profiles (\cite{atkinson2007optimum}).
To name a few, \cite{park2018optimal} employed a gradient-based algorithm for efficient parameter sensitivity analysis and optimal fitting. Optimized current profiles demonstrate accelerated parameter identification compared to standard methods. In the same way, \cite{sharma2014fisher} used FI to quantify the identifiability of battery parameters for a first-order nonlinear equivalent-circuit model subjected to periodic cycling. Similarly, \cite{lai2021new} introduced an OED strategy for an electrochemical battery model, focusing on enhancing the identifiability of diffusion parameters through parameter sensitivity analysis. Likewise, \cite{mendoza2017maximizing} examined the feasibility of a single current cycling experiment to identify the parameters of electrochemical and thermal models, and \cite{rothenberger2015genetic} conducted OED based on input shaping for maximizing identifiability using FI. 

While these studies collectively underscored the potentials of OED in advancing battery identifiability, it is crucial to acknowledge the inherent challenges in its practical applicability. First, OED techniques, despite their promises, often involve solving nonlinear and non-convex optimization problems in an offline fashion that are both time-consuming and resource-intensive, leading to practical implementation issues. Moreover, there exist notable concerns regarding the resulting unrealistic current profiles that might pose safety risks, such as overcharging, overheating, or battery failure (\cite{jaguemont2023critical}). This is largely due to aggressively pushing batteries beyond safe limits in search of a maximized FI based on assumed parameters that may not accurately represent the underlying variability and complexity of battery dynamics.

These drawbacks reverberate into the MPC performance, where the reliance on accurate models becomes a critical issue. A minor discrepancy between predicted and actual battery behavior, resulting from inaccurate parameter assumptions, directly impacts the efficacy of control strategies during charging. Additionally, the limited adaptability of fixed-parameter models hinders MPC's ability to respond effectively to evolving battery conditions.

Given these critical considerations, the transition from offline OED for battery identifiability to the online implementation of MPC for safe battery charging presents complex challenges that need to be addressed holistically. It is imperative to bridge the gap between offline design and online implementation to optimize battery charging.
To achieve this, enhancing battery model accuracy through refined identifiability methodologies in OED can serve as a robust foundation for effective MPC implementation; and by integrating real-time adaptation mechanisms within MPC, such as adaptive parameter estimation or dynamic model updating, the system can better accommodate the inherent variability and complexity of battery dynamics.

In this regard, we propose a novel approach that integrates OED into an adaptive MPC framework to safely fast charge a battery while enhancing parameter identifiability in real-time. By synthesizing these two concepts and addressing online computational challenges, we decompose the optimization problem into two distinct phases: (i) an offline optimization phase aimed at generating reference charging protocols with maximized parameter identifiability; and (ii) an online control phase that leverages pre-computed information for an adaptive charging control.
Specifically, the offline optimization phase focuses on maximizing FI while also fulfilling charging objectives with safety constraints. This phase produces an optimized state of charge (SOC) trajectory to be used in the online control phase. 
Subsequently, in the online control phase, 
the adaptive MPC will track the pre-optimized SOC trajectory from the offline optimization phase to calculate the real-time current. 
This strategy aims to replicate in real-time the optimal input current produced in the offline phase that maximizes FI. 
Notably, in this work, instead of simply applying the offline current for battery charging, MPC-based online control is essential for 
dynamically adjusting to changes related to uncertain model parameters (\cite{lin2014lumped}).
Hence, to enhance adaptability and robustness against parameter uncertainty, we will incorporate a real-time parameter updating scheme, which tracks model parameters to ensure that the predictive model can accurately respond to evolving conditions. 



To summarize, this work outlines the following technical contributions:
\vspace{0.2 cm}
\begin{itemize}
    \item Novel integration of Fisher information into an adaptive MPC for battery charging control, enhancing parameter identifiability and real-time adaptability.
    \item Emphasis on battery safe operation during charging and input shape flexibility in maximizing parameter identifiability.
    \item Inclusion of online parameter estimation within the MPC framework to improve model predictive accuracy and system robustness.
\end{itemize}
\vspace{0.2 cm}
To the authors' best knowledge, this paper marks the first attempt in the literature to integrate OED into a model predictive control framework for improving real-time battery safe charging while taking parameter identifiability into consideration. This work contributes to improving battery safety, adaptability, and predictive accuracy during battery charging operations.


\vspace{-0.1cm}
\section{Battery Equivalent Circuit Model}
\vspace{-0.1cm}

The equivalent circuit model (ECM) is a widely adopted battery representation. This model simplifies the intricate electrochemical reactions  into a network of passive electrical components, such as resistors, capacitors, and a voltage source (\cite{hu2012comparative}).
By capturing the essential electrochemical phenomena in a simplified yet effective manner, ECM facilitates real-time simulation and control implementation in practical applications (\cite{plett2015battery}).
The model incorporates a open circuit voltage (OCV), which varies with the SOC of the battery, an internal resistance ($R_0$) and a parallel RC pair consisting of a resistance ($R_1$) and a capacitance ($C_1$). The RC parallel circuit captures the transient behaviors resulting from charge-transfer reactions at the electrode interface, with the time constant given by $\tau = R_1C_1$ (\cite{hu2012comparative}). The dynamic equations governing the model are as follows:
\begin{align}
    \dot{z}(t) & = \frac{1}{Q}I(t)\label{eq:ecm_model_z}, \\
    \dot{Q_c}(t) & = -\frac{1}{R_1C_1}Q_c(t)+I(t), \\
     V(t) & = OCV(z(t))+\frac{1}{C_1}Q_c(t)+R_0I(t), \label{eq:ecm_model_vt}
\end{align}
where $z(t)$ represents the SOC of the battery, and $Q$ denotes the charge capacity. Symbol $Q_c(t)$ indicates the amount of charge stored in the parallel capacitor $C_1$. The input current is denoted by $I(t)$ (with charging associated with positive current values), and the output terminal voltage is represented by $V(t)$.


\section{Fisher Information-based Adaptive Battery Charging Controller}

\begin{figure}[t!]
    \centering
    \includegraphics[width=0.78\columnwidth]{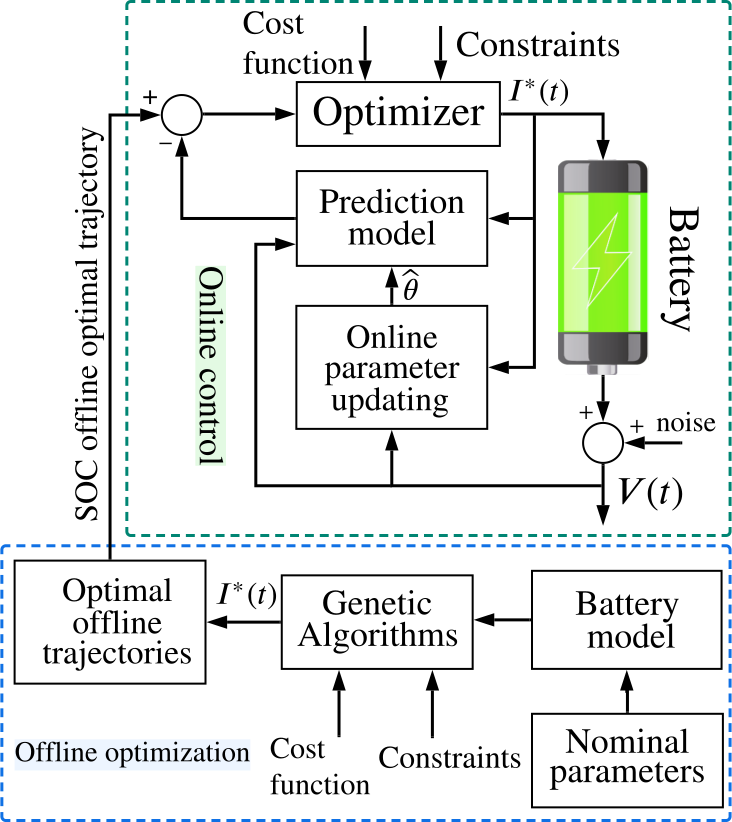}
    \caption{Proposed Fisher-information-guided adaptive battery controller scheme. }
    \label{fig:ctr_scheme}
\end{figure}

In this section, we delve into the design of the Fisher information-guided adaptive battery charging controller.  Fig.~\ref{fig:ctr_scheme} illustrates the proposed control scheme, which comprises two main phases: an offline optimization phase (the blue dashed-line block) and an online control phase (the green dashed-line block).

This approach leverages both offline optimization and online control techniques to maximize parameter identifiability and ensure safe battery operation during charging phase. In addition, the proposed control strategy allows for real-time parameter updating, see 
green dashed-line block in Fig. \ref{fig:ctr_scheme}, which is beneficial for adaptively optimizing battery charging process under evolving battery dynamics.




\subsection{Fisher Information for Parameter Estimation}

Battery parameter identification plays a pivotal role in accurately predicting battery behavior and in turn, enables the establishment of optimal and precise control policies that enhance battery performance. To maximize battery parameter identifiability, FI theory has become a widely explored field which offers a powerful toolset for analyzing the information content of experimental data about unknown parameters. This theory aids in battery parameter estimation by assessing the sensitivity of experimental voltage measurements regarding model parameters. In this context, a higher FI value leads to greater sensitivity, yielding that certain battery parameters might be estimated more precisely based on specific experimental data (\cite{fujita2022fisher}).
To quantify FI, a key mathematical concept known as Fisher information matrix (FIM) emerges. This metric provides a comprehensive measure of parameter sensitivity by evaluating how variations in model parameters can affect the likelihood of observed data. If the likelihood of data remains relatively unchanged regardless of variations in the parameter value, then data provide limited information about the parameter. Conversely, if the likelihood of data is highly sensitive, then data contain substantial information about the parameter, thereby enabling a more accurate parameter identification (\cite{chao2016fisher}).
Central to our study, we describe the battery model \eqref{eq:ecm_model_z}-\eqref{eq:ecm_model_vt} in a standard state-space representation:
\begin{align}
    \dot{x}(t)&= f(x(t),u(t),\theta),\\
    y(t)&= g(x(t),u(t),\theta)+v(t).
\end{align}
Here, $x(t)$ represents the internal states of the battery, $u(t)$ denotes the input current over time, $\theta$ describes the unknown battery parameters, and $y(t)$ is defined as the measured terminal voltage corrupted by a noise signal $v(t)$. Technically, the noise signal is inherently unknown; otherwise, it would be theoretically feasible to draw a perfect voltage signal by subtracting the noise from the measured voltage. Nevertheless, a crucial assumption in FI analysis relies on having a prior knowledge of the noise signal behavior (\cite{sharma2014fisher}).
Accordingly, the experimentally measured terminal voltage typically differs from the simulated values due to inherent output voltage noise. This enables the calculation of a likelihood function $p(y(t) |\theta)$, which quantifies the likelihood that the discrepancy between the battery's simulated output and the actual measured voltage is a realization of the assumed noise process (\cite{rothenberger2015genetic}, and \cite{sharma2014fisher}).

Building upon this likelihood function, the mathematical expression for FIM is defined as the expected value of the Hessian matrix of the log-likelihood function with respect to $\theta$, expressed as:
\begin{equation}\label{eq:FIM_E}
    \textbf{\text{FIM}}=\mathbb{E}\left\{\left(\frac{\partial}{\partial \theta} \ln{p(y(t)|\theta)}\right)\left(\frac{\partial}{\partial \theta} \ln{p(y(t)|\theta)}\right)^\intercal\right\}
\end{equation}
where, $\frac{\partial}{\partial \theta} \ln p(y(t) | \theta)$ represents the score function, offering insights into how changes in $\theta$ impact the log-likelihood function, while $\mathbb{E}$ denotes the expectation over the noise distribution. 

The battery model described by equations (\ref{eq:ecm_model_z})--(\ref{eq:ecm_model_vt}) uses a parameter set $\theta = \left[R_0, \frac{1}{C_1}, \frac{1}{Q}, \frac{1}{R_1C_1}\right]^\intercal$, yielding the estimated output voltage $V(t, \theta)$.

The parameterized battery model of equations (\ref{eq:ecm_model_z})--(\ref{eq:ecm_model_vt}) is expressed as follows:
\begin{align} 
    \dot{z}(t,\theta)&=\theta_3I(t),\\\label{eq:p_Qc} 
    \dot{Q_c}(t, \theta)&=-\theta_4Q_c(t,\theta)+I(t),\\   V(t,\theta)&=OCV(z(t,\theta))+\theta_2Q_c(t,\theta)+\theta_1I(t). 
\end{align}

For mathematical simplicity, assuming the noise signal is Gaussian and white, we derive the following FIM (\cite{sharma2014fisher}):

\vspace{-0.3cm}
\begin{equation}
    \textbf{\text{FIM}}=\frac{1}{\sigma^2}\int_0^T\left[\left( \frac{\partial V(t,\theta)}{\partial \theta}\right)\left( \frac{\partial V(t,\theta)}{\partial \theta}\right)^\intercal\right] dt.
\end{equation}

In the given expression, $\sigma$ represents the standard deviation of the measured terminal voltage. The sensitivity term $\frac{\partial V(t,\theta)}{\partial \theta_j}= S_{j}$, in which
$j = 1, 2, 3, 4$ indexes parameters, and $T$ denotes the period over which OED is conducted.

We now detail the derivations of the sensitivity terms for computing the FIM. The first two terms related to $\theta_1$ and $\theta_2$ are straightforward and can be expressed as:
\begin{align}
S_{1} &= \frac{\partial V(t,\theta)}{\partial \theta_1} = I(t), \\
S_{2} &= \frac{\partial V(t,\theta)}{\partial \theta_2} = Q_c(t,\theta).
\end{align}

To compute the sensitivity of the third term ($S_{3}$), we apply the chain rule:
\begin{equation}
S_{3} = \frac{\partial V(t,\theta)}{\partial \theta_3} = \frac{\partial V(t,\theta)}{\partial \text{OCV}(z(t,\theta))} \frac{\partial \text{OCV}(z(t,\theta))}{\partial z(t,\theta)} \frac{\partial z(t,\theta)}{\partial \theta_3}.
\end{equation}
The first partial derivative is equal to 1, and $\frac{\partial \text{OCV}(z(t,\theta))}{\partial z(t,\theta)} = \alpha(z(t,\theta))$, where $\alpha(\cdot)$ represents the slope of the OCV function. Next, to compute $\frac{\partial z(t,\theta)}{\partial \theta_3}$, given the initial condition $z(0) = z_0$, we use the relationship
\begin{equation}
z(t,\theta) = z_0 + \theta_3 \int_0^{t} u(\tau) d\tau.
\end{equation}

This yields the partial derivative with respect to $\theta_3$ as:
\begin{equation}
\frac{\partial z(t,\theta)}{\partial \theta_3} = \int_0^{t} u(\tau) d\tau.
\end{equation}
Considering the relation:
\begin{equation}
\int_0^{t} u(\tau) d\tau = \theta_3^{-1} (z(t,\theta) - z_0).
\end{equation}
Combining these expressions, we arrive at:
\begin{equation}
S_{3} = \alpha(z(t,\theta)) \theta_3^{-1} (z(t,\theta) - z_0).
\end{equation}

To obtain the last sensitivity term ($S_{4}$), we apply the chain rule, resulting in:
\begin{equation}
S_{4} = \frac{\partial V(t,\theta)}{\partial \theta_4} = \frac{\partial V(t,\theta)}{\partial Q_c(t, \theta)} \frac{\partial Q_c(t, \theta)}{\partial \theta_4}.
\end{equation}
It is straightforward to determine that $\frac{\partial V(t,\theta)}{\partial Q_c(t,\theta)} = \theta_2$, and we can handle the last partial derivative expression using the Laplace transform in equation (\ref{eq:p_Qc}) as follows:
\begin{equation}
Q_c(s,\theta) = \frac{1}{s+\theta_4} I(s).
\end{equation}
Next, we calculate the partial derivative with respect to $\theta_4$:
\begin{equation}
\frac{\partial Q_c(s,\theta)}{\partial \theta_4} = \frac{-1}{s^2 + 2\theta_4 s + \theta_4^2} I(s).
\end{equation}
The resulting transfer function $G(s,\theta)$ can be represented in the time domain by a linear state-space model using the controllable canonical form:
\begin{equation}
G(s,\theta) = \frac{-1}{s^2 + 2\theta_4 s + \theta_4^2}.
\end{equation}

\vspace{-0.1cm}
\subsection{Offline Optimization Problem Formulation}
\label{sec:offline}

\vspace{-0.1cm}
Within the scope of offline optimization problem formuation, the theory suggests that a larger FIM is preferred as it enhances the accuracy of parameter identification (\cite{rothenberger2015genetic}). To pursue this goal, we can maximize certain scalar metrics such as the matrix trace or its determinant. In our study, we focus on optimizing the determinant of the FIM. This approach is widely acknowledged in the literature as ``D-optimal" experimental design (\cite{atkinson2007optimum} and \cite{forman2013optimization}).  

We formulate our offline optimization problem as follows:

\vspace{-0.1cm}
\begin{mini}[3]
{I^{\text{off}}_{t_i},\dots, I^{\text{off}}_{t_N}}{-\text{det(\textbf{FIM})}}{}{}
\addConstraint{c_1:\text{battery dynamics in}~ (\ref{eq:ecm_model_z})-(\ref{eq:ecm_model_vt}),}{}
\addConstraint{c_2:z(t_f, \theta)}{=1,}{}
\addConstraint{c_3: I_{min}\le I^{\text{off}}(t_i)}{\le I_{max},\quad}{i=1,\dots,N.}
\addConstraint{c_4: V(t_i, \theta)}{\le V_{max},\quad}{i=1,\dots,N.}
\end{mini}

The objective of the offline optimization problem is to maximize the determinant of FIM to improve battery parameter identifiability during safe battery charging. However, it is crucial to acknowledge that previous studies (\cite{rothenberger2014maximizing} and \cite{sharma2014fisher}) often overlook critical safety constraints, such as battery overcharge (constrained by $c_2$ and $c_4$), which is essential for avoiding potential battery damage like explosions or premature aging.

In our approach, we consider controlled rapid charging within a user-defined time frame $[0,t_f]$, under constraint $c_2$, aiming to ensure that battery reaches full charge at a specific time $t_f$ while simultaneously maximizing FI. 
Additionally, our method also offers increased flexibility in charge current shaping, enabling a more dynamic charging profile beyond traditional input-shaping techniques (\cite{rothenberger2014maximizing}, \cite{rothenberger2015genetic} and \cite{mendoza2017maximizing}). This enhanced flexibility improves sensitivity in the observed data, contributing to more accurate parameter estimation.


Mathematically, maximizing the determinant of FIM can be considered as a non-convex optimization problem (\cite{rothenberger2015genetic} and \cite{de2016d}) because of the nonlinearity of OCV curve, leading to an unbounded current profile. To address this challenge, existing literature suggests incorporating  additional objectives in the cost function, such as including the l2-norm of the current signal, to ensure that the resulting current profile remains reasonable and bounded (\cite{rothenberger2014maximizing}), which may compromise the importance of maximizing FI. Conversely, our proposal mitigates this issue by establishing constrained charging defined by $c_2$ and bounding the optimal current profile using $c_3$. To navigate the non-convex landscape and avoid local minima, we utilize genetic algorithms employing crossover, mutation, and selection strategies. These methods efficiently explore the multi-modal cost function surface, enhancing the search for global optima (\cite{bajpai2010genetic}). 
Ultimately, this comprehensive offline method not only focuses on enhancing identifiability but also prioritizes battery safety and charging efficiency.

The resulting optimal trajectories from this offline optimization step are utilized in the subsequent online control to ensure safe and efficient battery operation.

\subsection{Online Charging Control Using Adaptive MPC} \label{sec:online}

This section explores adaptive MPC to ensure that battery operational constraints are met while performing real-time parameter identification. By leveraging the pre-optimized SOC trajectory from the offline optimization, this approach is specifically designed to enhance battery parameter identifiability and safety simultaneously during online operation.

Adaptive MPC (see the green dashed-line block in Fig.~\ref{fig:ctr_scheme}) is an advanced control strategy that optimizes control actions over a finite time horizon $H$ using a predictive model (\cite{schwenzer2021review}). In battery charging, MPC ensures SOC tracking while adhering to safety constraints to prevent battery damage (\cite{pozzi2023imitation}). Mathematically, the battery charging problem can be framed as a constrained optimization, as illustrated below:

\vspace{-0.3cm}
\begin{mini}[3]
{I^{\text{on}}_{t_{k+1}},\dots,I^{\text{on}}_{t_{k+H}}}{ \sum_{i=k+1}^{k+H} \| z_{\text{ref}}(t_i) - \hat{z}(t_i,\hat{\theta}) \|^2}{}{}
\addConstraint{d_1:\text{battery dynamics in}~ (\ref{eq:ecm_model_z})-(\ref{eq:ecm_model_vt}),}{}
\addConstraint{d_2: z(t_i,\hat{\theta}) \le}{1,}{i=k+1,\dots,k+H.}
\addConstraint{d_3: I_{min}\le I^{\text{on}}(t_i)}{\le I_{max,}\quad}{i=k+1,\dots,k+H.}
\addConstraint{d_4: V(t_i, \hat{\theta})}{\le V_{max},\quad}{i=k+1,\dots,k+H.}\label{eq:mpc}
\end{mini}
\vspace{-0.2 cm}

Notably, $z_{\text{ref}}(t_i)$ represents the pre-optimized SOC trajectory produced from the offline optimization phase in Section~\ref{sec:offline}. This pre-optimized SOC trajectory, as previously discussed, was generated by maximizing FI over the entire time horizon $[0,t_f]$. By tracking $z_{\text{ref}}(t_i)$ in \eqref{eq:mpc}, the solution produces a dynamical online charging current in an effort to mimic the current profile from the offline optimization so as to maximize FI and improve parameter estimation in real time. 
Different from other existing works, MPC in this paper incorporates real-time parameter updating. This methodology minimizes the discrepancy between actual battery voltage measurements and the corresponding estimated voltages by iteratively updating the estimated parameters $\hat{\theta}$ within specified bounds.  This adaptive approach improves model accuracy, allowing for the adaptation to varying battery conditions over time (e.g., aging) while still adhering to operational battery constraints. 

\textbf{Remark.}
\textit{Parametric uncertainties in OED studies can lead to sub-optimal trajectories when incorrect nominal model parameters are used during the offline design phase (\cite{rothenberger2014maximizing},~\cite{mendoza2017maximizing}, \cite{rojas2007robust}). Consequently, the online current profile computed by the adaptive MPC strategy might differ from the pre-optimized offline profile due to discrepancies between the true battery parameters and the nominal parameters assumed during the offline phase.
However, for this study, online parameter adaptation within the MPC framework mitigates the influence of parametric uncertainties by updating parameter values in real-time to compensate for model inaccuracies, maintaining optimality despite possible offline design deficiencies.}

\textit{Future work aims to enhance pre-optimized trajectory reliability by investigating advanced offline design techniques like Bayesian optimal design (\cite{rothenberger2015robust}), which incorporates prior parameter distribution statistics into the offline optimization design, potentially reducing the existing offline-online optimal profile discrepancy.}

\section{Simulation Results}
As part of this analysis, we present the results of our adaptive MPC approach utilizing Fisher information for real-time parameter identification during safe battery charging.

\subsection{Problem Setup}
As has been widely acknowledged
in the literature, a critical issue in OED
is that a set of nominal parameters must be presumed known \textit{a priori} for maximizing FI (\cite{rothenberger2014maximizing}, \cite{mendoza2017maximizing}, \cite{rojas2007robust}). In this work, the nominal battery parameters are given by
 Table \ref{tab:nom_param_ecm}. 
However, this assumption may not hold valid in practice, as actual battery parameters may deviate from the nominal values initially assumed. Therefore, we consider the following true parameters for the battery model: $R_{0,true}=0.05~ \Omega, ~C_{1,true}=950 ~F,~ Q_{true}=1.9~ \text{Ah}$, while $R_{1,true}=0.03 ~\Omega$. This adjustment enables us to demonstrate how the adaptive charging controller converges in real-time toward these true battery parameters to ensure the accuracy and reliability of the battery model over time.
Moreover, the nonlinear OCV function  used in this study is derived from experimental data provided by the Center for Advanced Life Cycle Engineering (\cite{Calce}), using 1/20 C-rate on a Samsung INR-18650 20R cell. The OCV curve is modeled using a 7th-order polynomial function 
with respective coefficients ($a_0,\dots,a_7$) given in Table \ref{tab:nom_param_ecm}.

\begin{table}[th!]
    \centering
    
    \caption{Nominal parameters and coefficients of OCV curve}
    \vspace{-0.15cm}
    \begin{tabular}[width=0.1\textwidth]{cccc}
        \toprule[0.3mm]
        \textbf{Parameter} & \textbf{Value} & \textbf{Parameter} & \textbf{Value}\\
        \midrule[0.3mm]
        $Q$&	2.0 Ah&	$a_2$&- 77.31237098\\
        $R_1$&	0.03 $\Omega$&	$a_3$&327.446181\\
        $C_1$ &	1000 $F$&	$a_4$&- 763.3324119\\
        $R_0$ &	0.06 $\Omega$&	$a_5$&988.408671\\
        $a_0$& 3.039475779&	$a_6$&- 662.9843922\\
        $a_1$& 9.620312047&	$a_7$&179.301862\\
         \bottomrule[0.3mm]
    \end{tabular}
    \label{tab:nom_param_ecm}
\end{table}

Key considerations to highlight include the following: 
We are implementing a constrained fast charging protocol aimed at achieving a full battery charge within 30 minutes ($t_f = 1800$ seconds). The dynamic charging current is constrained to -3C to 3C, with a voltage limit of 4.3 V to prevent overcharging.
We employ a GA with a population size of 50 individuals. The optimization is conducted on a high-performance computer with an Intel Core i9 processor with 64 GB of RAM and 24 cores, also leveraging parallel processing to expedite solution convergence within the MATLAB software environment. 
Furthermore, the adaptive MPC utilizes \texttt{fmincon} to solve the optimization problem. The same computational processor and software environment are utilized.

\subsection{Discussion and Future Work}
\begin{figure}[t!]
    \centering
    \includegraphics[width=1\columnwidth]{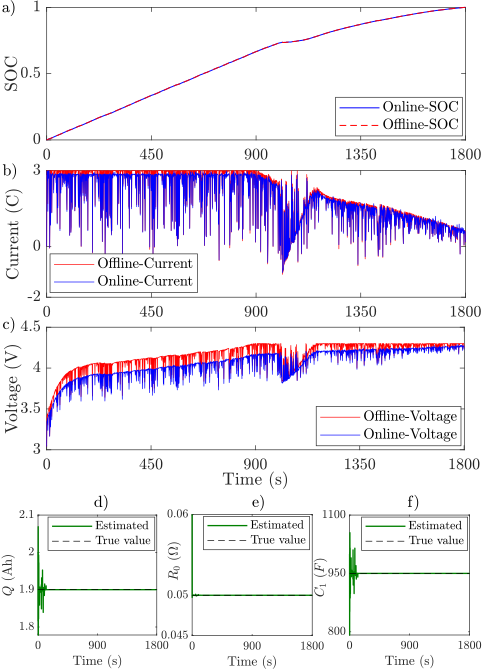}
    \vspace{-0.5cm}
    \caption{Comparison of offline optimized trajectories with online adaptive MPC responses for battery charging and real-time parameter updating.}
    \label{fig:off-vs-on}
\end{figure}


Our results in Fig. \ref{fig:off-vs-on}.a) demonstrate the controller's capability to safely charge the battery within a 30-minute interval, maximizing the FI metric for real-time parameter estimation. The high tracking accuracy depicted in Fig. \ref{fig:off-vs-on}.a) significantly influences the observed discrepancy in Fig. \ref{fig:off-vs-on}.c), attributed to practical differences in battery model parameters between offline optimization and online control. Throughout the charging period, Fig. \ref{fig:off-vs-on}.c) confirms consistent adherence to safety constraints, including the maximum voltage limit, ensuring battery operation remains within a safe zone to prevent potential battery damage.


Fig. \ref{fig:off-vs-on}.b) shows the difference between the offline optimized current profile and the online adaptive MPC current profile due to the discrepancy between the assumed nominal battery parameters in the offline optimization and the true parameter values used online. This discrepancy is further evident from the contrast in the determinant of the FIM calculated offline ($1.6275 \times 10^{40}$) and online ($3.8596 \times 10^{39}$) using the true parameter values. The smaller online FI value is expected due to practical variability in battery parameters during real-time operation. However, the adaptability of online adaptive MPC, which dynamically adjusts to evolving parameter values, may compensate for the reduced FI value, leading to improved parameter estimation accuracy over time, demonstrating the method's practical utility in refining model estimates and enhancing system performance despite parameter variability.

The proposed FI-based controller demonstrates its capability to accurately identify and converge towards the true battery model parameters in real-time, as illustrated in Figs. \ref{fig:off-vs-on}.d), \ref{fig:off-vs-on}.e), and \ref{fig:off-vs-on}.f). This real-time parameter updating feature contributes to enhancing the reliability and accuracy of the battery model over time, ensuring that the controller can effectively mitigate parametric uncertainties by adapting to evolving conditions.

Future directions of our work highlight the following insights: (i) accounting for the dependency of battery model parameters on temperature, SOC, or aging; (ii) enhancing the offline optimization phase by incorporating more advanced optimal design techniques, rather than relying solely on nominal values, to better capture the underlying parameter variability and uncertainty; (iii) transitioning towards more sophisticated battery models, such as physics-based models, to leverage additional features of this approach, including the consideration of battery degradation during charging; and (iv) validating our method with experimental data to demonstrate its robustness in a more demanding environment.

\section{Conclusions}

This manuscript explores a pioneering solution of integrating FI into an online control framework for safe battery charging. With the objective of minimizing computational complexity we consider partitioning the optimization process into offline and online phases. The findings illustrate a constrained safe charging protocol capable of maximizing FI in real time. This underscores the viability of this approach for real-time parameter identification in Li-ion batteries characterized by inherently poor parameter identifiability. Future directions for this research involve transitioning towards more sophisticated electrochemical models to capture detailed battery behavior and conducting experimental validations to showcase the effectiveness of Fisher Information-guided adaptive battery controller.

\section*{Acknowledgement}

This material is based upon work supported by the National Science Foundation (NSF) under Award No. 2327327. Any opinions, findings, and conclusions or recommendations in this paper are those of the author(s) and do not necessarily reflect the views of the NSF.

\bibliography{ifacconf}             
                                                   







\end{document}